\documentclass{elsart}
\usepackage{graphicx}
\usepackage{hyperref}
\usepackage{natbib}
\usepackage{amsmath}

\journal{Theoretical Population Biology}

\begin{document}

\title{Effect of Predators of Juvenile Rodents on the Spread of the
Hantavirus Epidemic}

\begin{frontmatter}

\author[consortium]{G. Camelo-Neto}, \ead{gustavo@phys.unm.edu}
\author[consortium,ana]{Ana T.C. Silva}, \ead{tereza@uefs.br}
\author[consortium]{L. Giuggioli}, \ead{giuggiol@unm.edu}
\author[consortium]{V.M. Kenkre} \ead{kenkre@unm.edu}

\address[consortium]{Consortium of the Americas for Interdisciplinary  
Science and Department of Physics, University of New Mexico,
Albuquerque, NM 87131, USA.}

\address[ana]{Departamento de F\'isica, Universidade Estadual de Feira de Santana,
Feira de Santana, BA 44031-460, Brazil}

\date{\today}

\begin{abstract} Effects of predators of juvenile mice on the spread of
the Hantavirus are analyzed in the context of a recently proposed model.
Two critical values of the predation probability are identified. When
the smaller of them is exceeded, the hantavirus infection vanishes
without extinguishing the mice population. When the larger is exceeded,
the entire mice population vanishes. These results suggest the
possibility of control of the spread of the epidemic by introducing
predators in areas of mice colonies in a suitable way so that such control
does not kill all the mice but lowers the epidemic spread.
\end{abstract}



\end{frontmatter}

\section{Introduction and the Model}
\label{intoroduction}

The Hantavirus epidemic~\citep{yatesreview,mills99}, known to be spread
by the movement of, and virus transmission between, mice, has received a
lot of mathematical attention in the last few
years~\citep{AK,aguirre02,abramson03,vmkpasi,buceta04,escudero04,vmkphysicaA05}. Recently, we
proposed a model~\citep{liqsolid} characterized by two types of
mice: itinerant juvenile mice known technically in the biological
literature as subadults  that roam in their attempts to find their own
home ranges until they find a suitable place, at which time they turn
into adult mice; and the relatively less mobile adult mice that restrict
their movements within their own home ranges. Given the two extreme
types of movement (freely  diffusing and static) that the juveniles and
the adult perform in the field, the model has been termed the
\emph{liquid-solid} (LS) model. An analysis of the LS model has been
provided~\citep{liqsolid} on the basis of mean field calculations as well
as computer simulations that incorporate spatial features. To simplify
the analysis it was assumed in that work that the juvenile
mice do not die; that they cease to exist only when they grow into adult
mice on finding a suitable site to call their own home range. While it
allows a quicker insight into the consequences of the main aspects of
the model, that assumption is surely not realistic in many situations.
The itinerant juvenile mice might meet with predators in the open field
and be killed by them. The purpose of the
present study is to explore the effects of the predator-induced
attrition in the juvenile population.

Our analysis below shows that two transitions occur as a direct
consequence of this realistic feature. The infected phase vanishes for
sufficiently large predation `pressure'  whatever the value of the other
parameters. In particular, this happens even in the presence of
unlimited environment resources (e.g., food). The second transition is
the appearance of a bifurcation when the rate at which the juveniles are
converted into adults is varied. A population is sustained by the
environment only if juveniles grow into adults fast enough to avoid
being killed by the predators.

As explained in our earlier analysis~\citep{liqsolid}, the LS model may be represented at
the \emph{kinetic} level by
\begin{eqnarray}
\frac{\partial B_{i}(x,t)}{\partial t} &=&-c_{B}B_{i}-\frac{B_{i}(A+B)}{K(x,t)}+aB_{s}(A_{i}+B_{i})+D\nabla ^{2}B_{i}-G(x)B_{i},  \nonumber \\
\frac{\partial B_{s}(x,t)}{\partial t} &=&bA-c_{B}B_{s}-\frac{B_{s}(A+B)}{K(x,t)}-aB_{s}(A_{i}+B_{i})+D\nabla^{2}B_{s}-G(x)B_{s},  \nonumber \\
\frac{\partial A_{i}(x,t)}{\partial t} &=&-cA_{i}-\frac{A_{i}(A+B)}{K(x,t)} +aA_{s}B_{i}+G(x)B_{i}, \nonumber \\
\frac{\partial A_{s}(x,t)}{\partial t} &=&-cA_{s}-\frac{A_{s}(A+B)}{K(x,t)} -aA_{s}B_{i}+G(x)B_{s}, 
\label{lsmodel}
\end{eqnarray}
where $A$ and $B$ (without suffixes)
denote the total densities of the adult and juvenile mice respectively,
the suffixes $i$ and $s$ represent infected and susceptible states of
the mice, and the last terms in each equation describe the settling down
of the juveniles into their own homes, accompanied by their conversion
into (static) adults. The rate of such conversion, $G(x)$, is non-zero
only if the spatial coordinate of the mouse, $x$, lies in regions that
the juveniles find suitable as their home ranges. An extreme way of
representing the confinement of the adults to their home ranges, used in
this LS model, is to take the adults to be immobile. Because we use such
a representation for simplicity~\citep{liqsolid},  there are no spatial
derivatives in the equations for the adults. The other parameters of the
model are as follows: $b$ is the birth rate of the juveniles, $c$ is the
death rate of the adults, $a$ is the infection parameter, $D$ is the
juvenile diffusion constant, and $K$ is the environment parameter  that,
for the sake of simplicity in this analysis, we consider to be
independent of time and space. Our particular focus is on the
introduction of $c_{B}$, the death rate of the juveniles. That
parameter, assumed zero for simplicity in our earlier analysis, is here
controlled by the presence of the predators.

\section{Simulation analysis}
\label{simul}

Our analysis here is at the configuration
master equation approach rather than at the kinetic level implied by
(\ref{lsmodel}). For this purpose we carry out simulations on a $L\times
L$ square lattice with each site of the lattice corresponding to a small
region in the landscape. We use moderately large lattices (with a total
of $2^{14}$ sites) and discrete time evolution. At each time step, the
juveniles may move but the adults not, the probability for the diffusive
(random walk) motion being 0.125 for any of the 8 directions of the
square lattice. Our time step is scaled in this manner to the diffusion
constant. An adult, infected or susceptible, gives birth to a
susceptible juvenile with probability $P_{b}$. An adult dies by aging
with probability $P_{c}$. When two or more mice meet at a site, they
compete and one may die with probability $1-P_{K}$. If a susceptible
mouse occupies the same site as an infected mouse, the former has
probability $P_{a}$ of getting infected in the next time step. A
juvenile dies because of predation with probability $P_{p}$ and, if it is
at a site without an adult, grows up settling in that site with
probability $P_{g}$. Corresponding to the kinetic level parameters of
equations (\ref{lsmodel}), $b$, $c$, $c_{B}$, $g$, $a$, and $K$, we have
the respective probabilities in the simulation description, $P_{b}$,
$P_{c}$, $P_{p}$, $P_{g}$, $P_{a}$, and $P_{K}$. Our focus in the
present study is on $P_p$.
\begin{figure}[tbh]
\centering
\includegraphics*[width=\columnwidth]{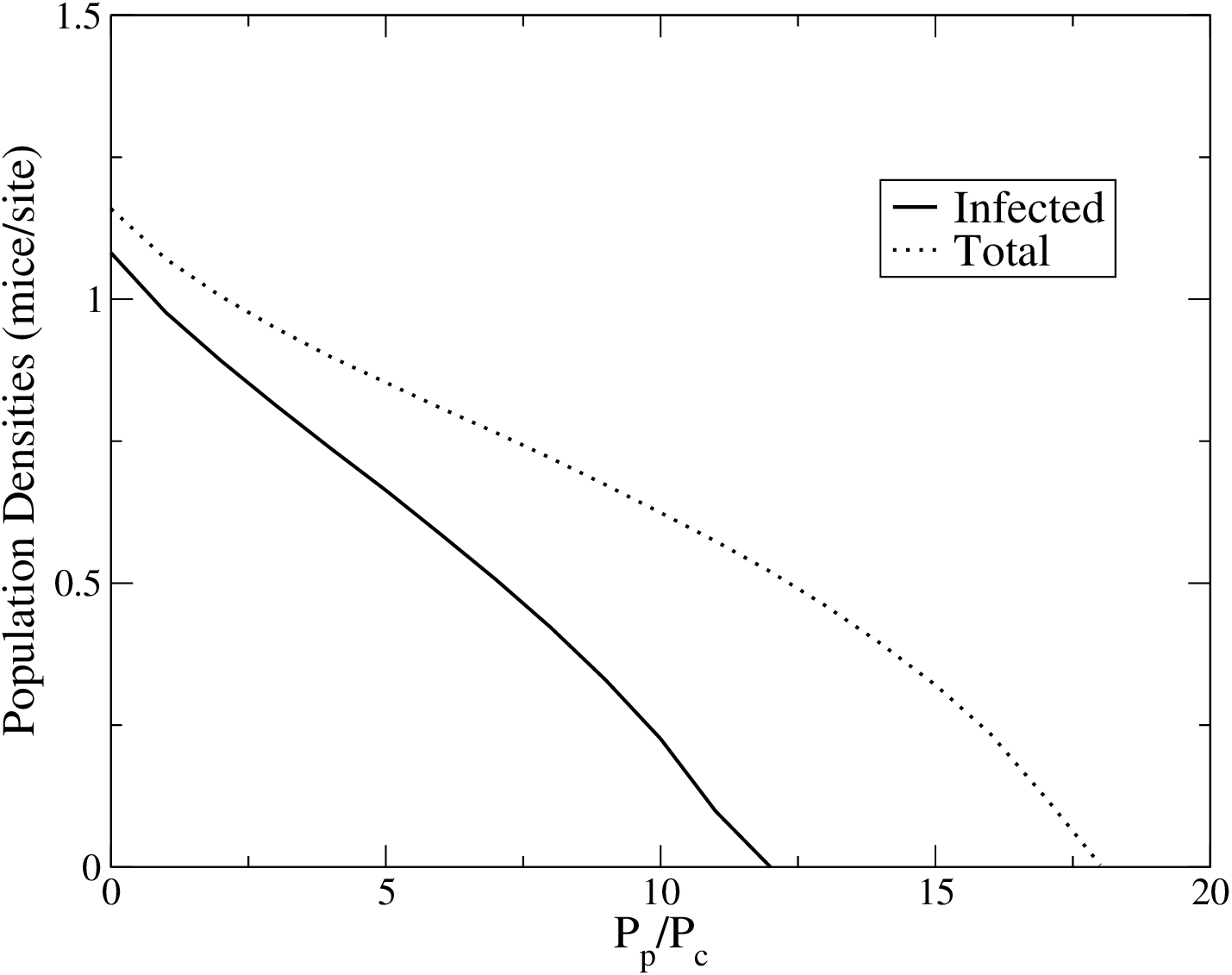}
\caption{Steady state densities of the infected and susceptible
populations versus $P_{p}/P_{c}$, the probability of the
predation-induced death of the juveniles divided
by the death probability of the adults.
Other parameters are  $P_{c}=0.01$, $P_{b}=0.02$, $P_{g}=0.6$,
$P_{a}=0.3$ and $P_{K}=0.98$. }
\label{ssPp}
\end{figure}

Our extensive computer simulations with attention on effects that are
qualitatively different from the case of vanishing $P_{p}$ analyzed
earlier, show a clear existence of separate infected, non-infected and
no-life phases. In Fig. \ref{ssPp} the steady state values of the
infected and the total mice populations are plotted as functions of the
ratio of the predation probability $P_{p}$ to the adult death
probability $P_c$. Sufficiently large values of the ratio make the
total mice population disappear because the predators kill all the
juveniles and the adults die their natural deaths. This is trivially
expected. However, an important finding in Fig. \ref{ssPp} is related to
the existence of the middle region of the $x$-axis, lying between the
values of $12$ and $18$ for the particular parameter values chosen,
where the infected population has died out but the total population has
not. The fact that predation thus buffers the transmission of infection
nontrivially, and may be thus used to eliminate it without killing off
all mice, is worthwhile to note.
\begin{figure}[tbh]
\centering
\includegraphics*[width=\columnwidth]{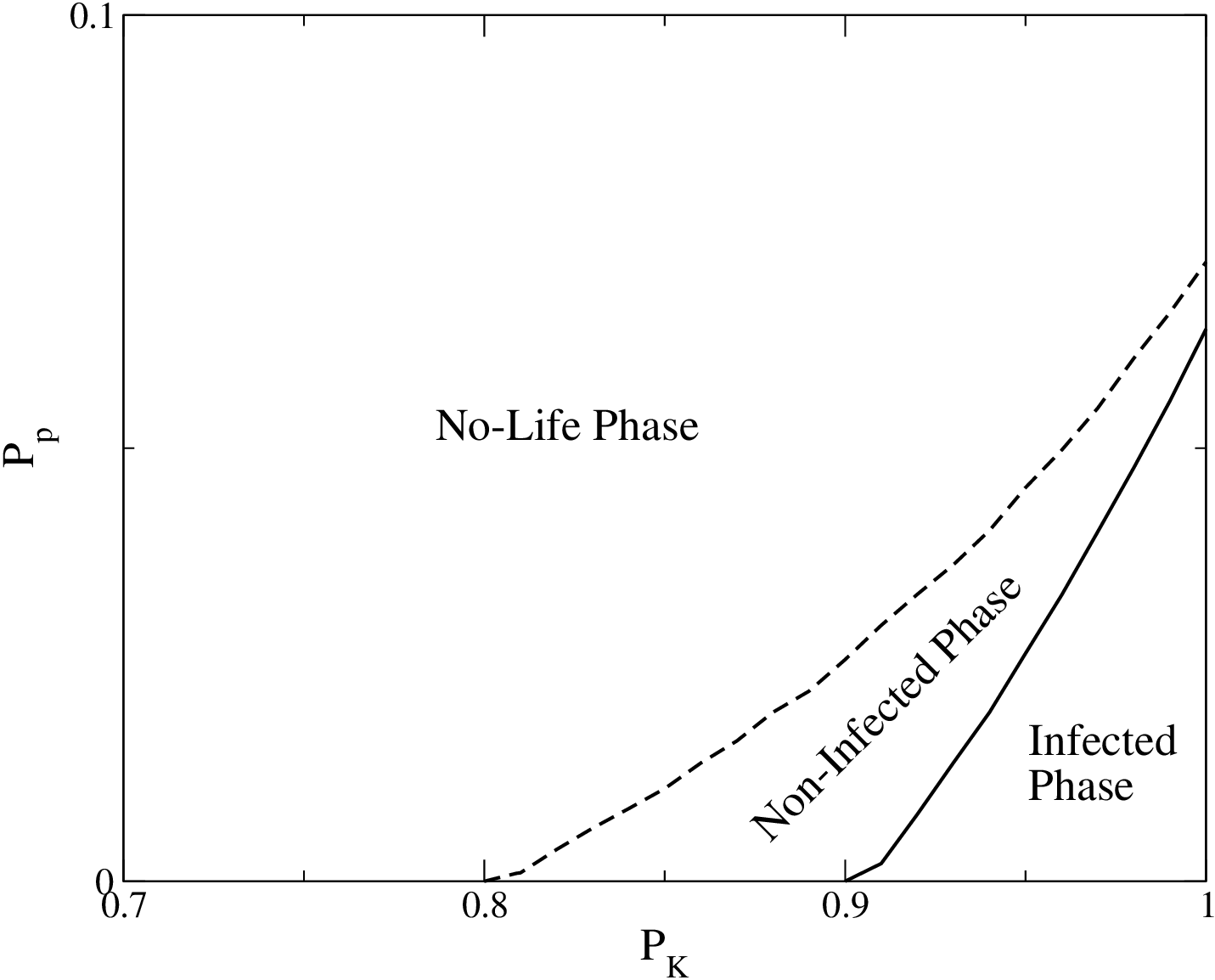}
\caption{$P_{p}-P_{K}$ Phase plot showing system behavior in various
regions of the predation probability $P_{p}$ and the
environment-controlled survival probability $P_K$. There are clear
regions where there is an infected phase, a non-infected phase, and a
phase in which no mice are living. This shows that if the predator
`pressure' is in the right level, infection can be eliminated without
killing all mice. The other parameters are: $P_{c}=0.01$, $P_{b}=0.02$,
$P_{g}=0.1$ and $P_{a}=0.3$.}
\label{PpPksimul}
\end{figure}

One characteristic of the model (\ref{lsmodel}), for the case $P_{p}=0$, is the
fact that, as competition decreases, eventually an infected phase appears
(if $P_{b}>P_{c}$ and irrespective of $P_{a}$ and $P_{g}$).
In other words, as the environment accommodates a larger
population, infection will always appear beyond a certain value of
$P_{K}$. This does not necessarily happen when $P_{p}\neq 0$. The
$P_{p}-P_{K}$ phase plot in Fig. \ref {PpPksimul} makes this clear. For a
certain range of $P_{p}$ values, the system remains susceptible for any
value of $P_{K}$. An increase in food availability has only the effect
of increasing the susceptible populations. Thus, the
introduction of an appropriate number of predators in an infected area
may completely eliminate the infection.
It is the presence of clusters of adults in the landscape that is
responsible for this effect.
If $P_{p}=0$, a juvenile cannot find a free site where to settle inside
a cluster and is forced to roam for a long period of time, increasing the chance
of being infected and transmitting the infection.
On the other hand, when $P_{p}>0$, the
adult clusters reduce the transmission of infection because the juveniles
are subject to predation as long as they roam in the landscape and do
not settle into a home range. A susceptible population may thus exist
without infection. Since any increment in the environmental resources
reduces competition and increases the cluster size, if
$P_{p}$ is sufficiently large, an increase of  $P_{K}$
maintains the entire population without infection.
\begin{figure}[tbh]
\centering
\includegraphics*[width=\columnwidth]{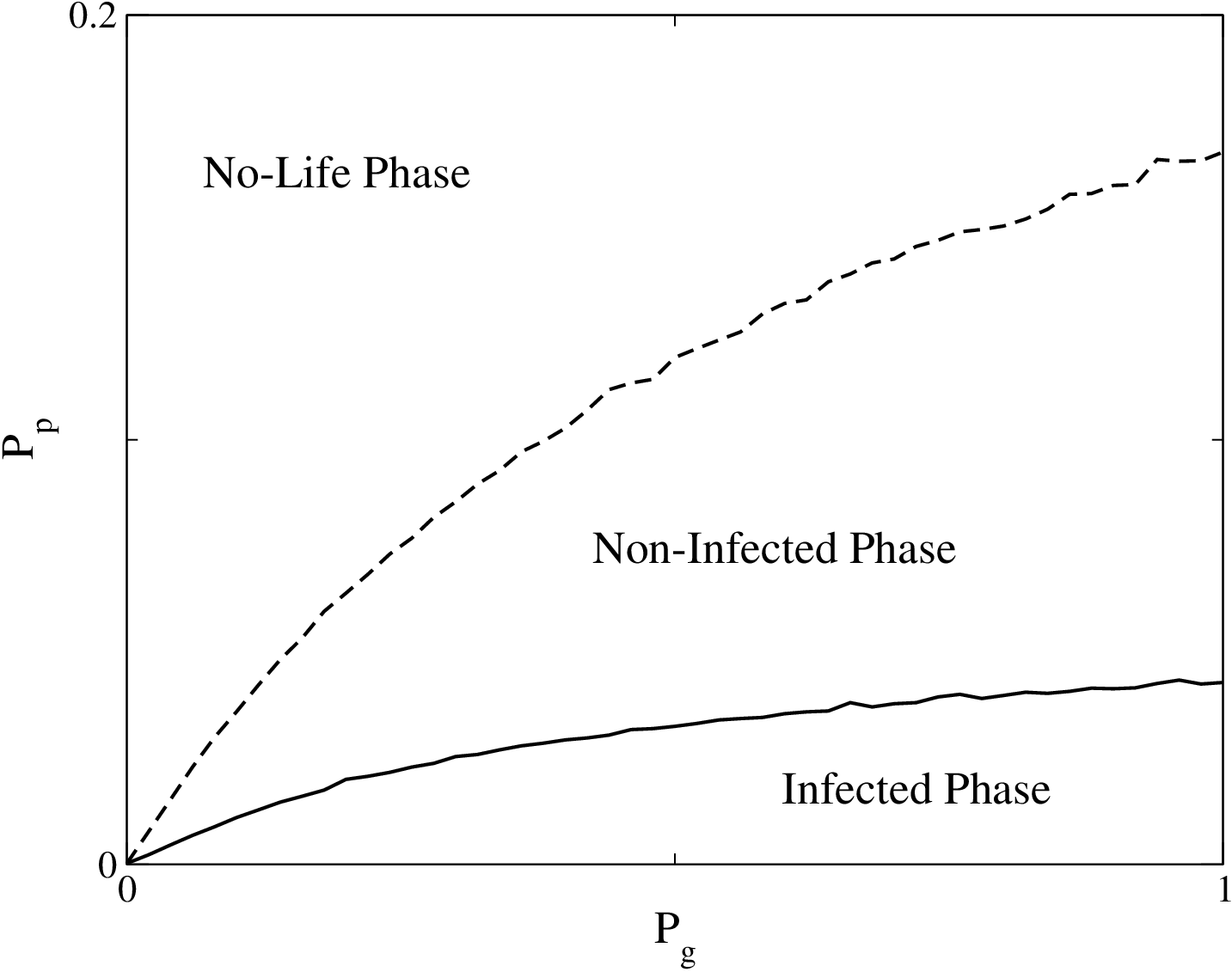}
\caption{$P_{p}-P_{g}$ Phase plot showing system behavior in various
regions of the predation probability $P_{p}$ and the probability $P_{g}$
at which juveniles grow into adults. Note how the parameter region where
the infection is absent (because of predation) gets larger as  $P_{g}$
increases. See text for explanation. The remaining parameters are 
$P_{c}=0.01$, $P_{b}=0.02$, $P_{K}=0.925$ and $P_{a}=0.3$.}
\label{PpPgsimul}
\end{figure}

The dependence of this infection-free regime on the other parameters is
studied in Fig. \ref{PpPgsimul} where the $P_{p}-P_{g}$ phase plot is
shown. In particular, we see how the window of $P_{p}$ values, where no
infection appears, gets larger as $P_{g}$ increases.
Bigger values of $P_{g}$ reduces the effect of predation by converting the juveniles
more rapidly into adults. In order to reach the condition with zero steady state
population, a larger $P_{p}$ value is necessary.
From Fig. \ref{PpPgsimul} we also see a different qualitative characteristic of the steady
state populations depending on whether $P_{p}$ is zero or larger than zero.
For a given $P_{p}>0$, the system may possess two bifurcations as $P_{g}$ is varied.
The first is related to the existence of a non-zero susceptible
population, and the second to the appearance of the infected phase.
On the other hand, if $P_{p}=0$, these two transition points merge into
a single one at $P_{g}=0$.
In order to better understand the existence of these bifurcations as
$P_{g}$ is varied, we approach the system through a
mean field analysis in the next section.

\section{Understanding the results from mean field considerations}
\label{mf}

An understanding of the simulations results reported above may be
attempted through mean field considerations. A mean field (no space
resolution) representation of the model in equations (\ref{lsmodel}) above,
equivalently in the simulations rules set out in Sec. \ref{simul}, is
\begin{eqnarray}
\frac{d\mathcal{B}_{i}}{d\tau } &=&-(\gamma+\delta) \mathcal{B}_{i}+\alpha %
\mathcal{B}_{s}\mathcal{A}_{i}+2\alpha
\mathcal{B}_{i}\mathcal{B}_{s}-%
\mathcal{B}_{i}\left( \mathcal{A}+\mathcal{B}\right) ,  \nonumber \\
\frac{d\mathcal{B}_{s}}{d\tau } &=&-(\gamma+\delta)\mathcal{B}_{s}+\beta \left(
\mathcal{A}_{s}+\mathcal{A}_{i}\right) -\alpha
\mathcal{B}_{s}\mathcal{A}%
_{i}-2\alpha\mathcal{B}_{i}\mathcal{B}_{s}-\mathcal{B}_{s}\left(
\mathcal{A}+\mathcal{B}\right),  \nonumber \\
\frac{d\mathcal{A}_{i}}{d\tau } &=&-\mathcal{A}_{i}+\alpha
\mathcal{A}%
_{s}\mathcal{B}_{i}+\gamma \mathcal{B}_{i}-\mathcal{A}_{i}\left(
\mathcal{A}+%
\mathcal{B}\right) ,  \nonumber \\
\frac{d\mathcal{A}_{s}}{d\tau } &=&-\mathcal{A}_{s}-\alpha
\mathcal{A}%
_{s}\mathcal{B}_{i}+\gamma \mathcal{B}_{s}-\mathcal{A}_{s}\left(
\mathcal{A}+%
\mathcal{B}\right),
\label{meanfieldeq}
\end{eqnarray}
where $\tau =ct$. Each script character in (\ref{meanfieldeq}) is dimensionless and denotes
the ratio of the quantity described by the corresponding Roman character in (\ref{lsmodel})
and $cK$. The other dimensionless quantities are the parameters 
$\gamma =g/c$, $\beta =b/c$, $\delta=c_{B}/c$ and $\alpha=Ka$.
As explained
elsewhere~\citep{liqsolid}, we take the rate associated with the
transfer of infection between two juveniles to be twice that between a
juvenile and an adult. This is in
keeping with similar considerations of the description of excitation capture and
annihilation processes in molecular
crystals~\citep{kenkreGMEbook,popeswenbergbook} and represents the situation
when motion of the juveniles is slow with respect to the actual infection
process.

The steady state solutions of the set of equations (\ref{meanfieldeq}) can
be found analytically in terms of the system parameters by noticing
that the total adult and juvenile population evolves in time according to
\begin{eqnarray}
\frac{d\mathcal{B}}{d\tau } &=&\beta \mathcal{A}-(\gamma +\delta
)\mathcal{B}-\frac{\mathcal{B}(\mathcal{A}+\mathcal{B})}{cK},  \nonumber \\
\frac{d\mathcal{A}}{d\tau } &=&-\mathcal{A}+\gamma
\mathcal{B}-\frac{\mathcal{A}(\mathcal{A}+\mathcal{B})}{cK}.
\label{sumeq}
\end{eqnarray}
It is easy to see from Eq. (\ref{sumeq}) that the total mice population
$\overline{\mathcal{A}}+\overline{\mathcal{B}}$
does not follow a logistic growth, contrary to what occurs in the original
model~\citep{AK} for the spread of the Hantavirus.
As a consequence,
$\overline{\mathcal{A}}+\overline{\mathcal{B}}$ is not linearly proportional to
$\beta $. It is straighforward to find from Eq. (\ref{sumeq})
the total population at steady state, i.e., the carrying capacity of the system
normalized to $cK$.
When $\beta <1+\delta /\gamma$, there is the trivial case
$\overline{\mathcal{A}}+\overline{\mathcal{B}}=0$, while,
with $\beta >1+\delta /\gamma $, the normalized carrying capacity is given by
\begin{equation}
\overline{\mathcal{A}}+\overline{\mathcal{B}}=\left( \beta -1-\delta
/\gamma \right) \frac{\gamma }{1+\delta +\gamma }\frac{\sqrt{1+2\chi}-1}{\chi },
\label{carcap}
\end{equation}
where $\chi =2(\beta-1-\delta /\gamma )\gamma (1+\gamma +\delta )^{-2}$.
Limiting analysis of Eq. (\ref{carcap}) shows that the carrying capacity increases linearly
with $\beta$ close to the bifurcation, and is proportional to
$\sqrt{\beta}$ for $\beta\rightarrow \infty$.

A stability analysis shows that the bifurcation at $\beta =1+\delta
/\gamma $ is transcritical. The interesting $\delta/\gamma$ dependence of this
bifurcation can be explained as follows. If juveniles die too fast
(large $\delta $) or become adult too slowly ($\gamma $ small),
further generation of juveniles is prevented. Both these mechanisms act
towards a reduction, and eventual disappearance, of the adult population.
The disappearance of the adult population, in turn, can make the entire population collapse to
zero at long times. The growth rate $\gamma $
counteracts the effect of $\delta $ because the conversion of juveniles to
adult status makes them resident and thereby inaccessible to predators in our
model. This dependence on $\delta /\gamma $ is reflected also in the
adult and juvenile population as can be seen from their values at steady
state.
\begin{eqnarray}
\overline{\mathcal{B}} &=&\left[ \frac{\left( \beta
-1-\delta /\gamma \right) }{
-1)}\right] \gamma \frac{(\beta +\delta +\gamma )\left( \sqrt{1+2\chi
}-1\right) -\chi (\gamma +\delta +1)}{\chi },\\ \nonumber
\overline{\mathcal{A}}&=&\left[ \frac{(1+\gamma )\left( \beta -1-\delta /\gamma \right)
}{(1+\gamma +\delta )(\beta +\delta -1)}\right] \gamma \frac{1+\chi
(1+\gamma +\delta )(1+\gamma )^{-1}-\sqrt{1+2\chi }}{\chi }.
\label{ssAB}
\end{eqnarray}

In the limit $\delta \rightarrow 0$, Eq. (\ref{ssAB}) reduces to mean field results
previously reported~\citep{liqsolid} in the absence of
predators. Some additional
algebra allows us to express the steady-state infected and susceptible
populations in terms of $\overline{\mathcal{A}}$ and
$\overline{\mathcal{B}}$. The phase
with no infection is simply given by
$\overline{\mathcal{B}}_{i}=0=\overline{\mathcal{A}}_{i}$,
$\overline{\mathcal{B}}_{s}=\overline{\mathcal{B}}$,
$\overline{\mathcal{A}}_{s}=\overline{\mathcal{A}}$, while 
the infected phase can be written as
\begin{equation}
\begin{array}{ll}
\overline{\mathcal{B}}_{i}=\frac{\sqrt{\mathcal{F}^{2}+8\mathcal{E}}-%
\mathcal{F}}{4\alpha }, & \qquad \overline{\mathcal{B}}_{s}=\overline{\mathcal{B}}-%
\overline{\mathcal{B}}_{i}, \\ 
\overline{\mathcal{A}}_{i}=\frac{\overline{\mathcal{B}}_{i}\left( \gamma
+\alpha \overline{\mathcal{A}}\right) }{\alpha \overline{\mathcal{B}}_{i}+1+%
\overline{\mathcal{A}}+\overline{\mathcal{B}}}, & \qquad \overline{\mathcal{A}}_{s}=%
\overline{\mathcal{A}}-\overline{\mathcal{A}}_{i},
\end{array}
\label{ssinf}
\end{equation}
wherein
$\mathcal{E}=\alpha \overline{\mathcal{B}}\left( \gamma +\alpha
\overline{\mathcal{A}}\right) -\left[ \gamma+\delta +\overline{\mathcal{A}}+
\overline{\mathcal{B}}\left( 1-2\alpha\right) \right] \left( 1+
\overline{\mathcal{A}}+\overline{\mathcal{B}}\right)$ and
$\mathcal{F} = 2(\gamma +1)+\overline{\mathcal{A}}\left( 3+\alpha\right) +\overline{\mathcal{B}}\left( 3-2\alpha\right)$.
Similarly to the mean field analysis~\citep{liqsolid} of this model in the absence of predators,
a stability analysis allows us to determine that infection
exists only when the infection parameter is larger than a critical value,
or similarly, when the environment parameter $K$ is larger than the
critical value:
\begin{equation}
K_{c}=\frac{\gamma +2\left(
1+\overline{\mathcal{A}}+\overline{\mathcal{B}}%
\right) }{2a\overline{\mathcal{A}}}\left\{
\sqrt{1+\frac{4\overline{%
\mathcal{A}}\left( \gamma+\delta
+\overline{\mathcal{A}}+\overline{\mathcal{B}}%
\right) \left( 1+\overline{\mathcal{A}}+\overline{\mathcal{B}}\right)
}{%
\overline{\mathcal{B}}\left[ \gamma +2\left( 1+\overline{\mathcal{A}}+%
\overline{\mathcal{B}}\right) \right] ^{2}}}-1\right\}.
\label{kcrit}
\end{equation}
\begin{figure}[tbh]
\centering
\resizebox{\columnwidth}{!}{\includegraphics{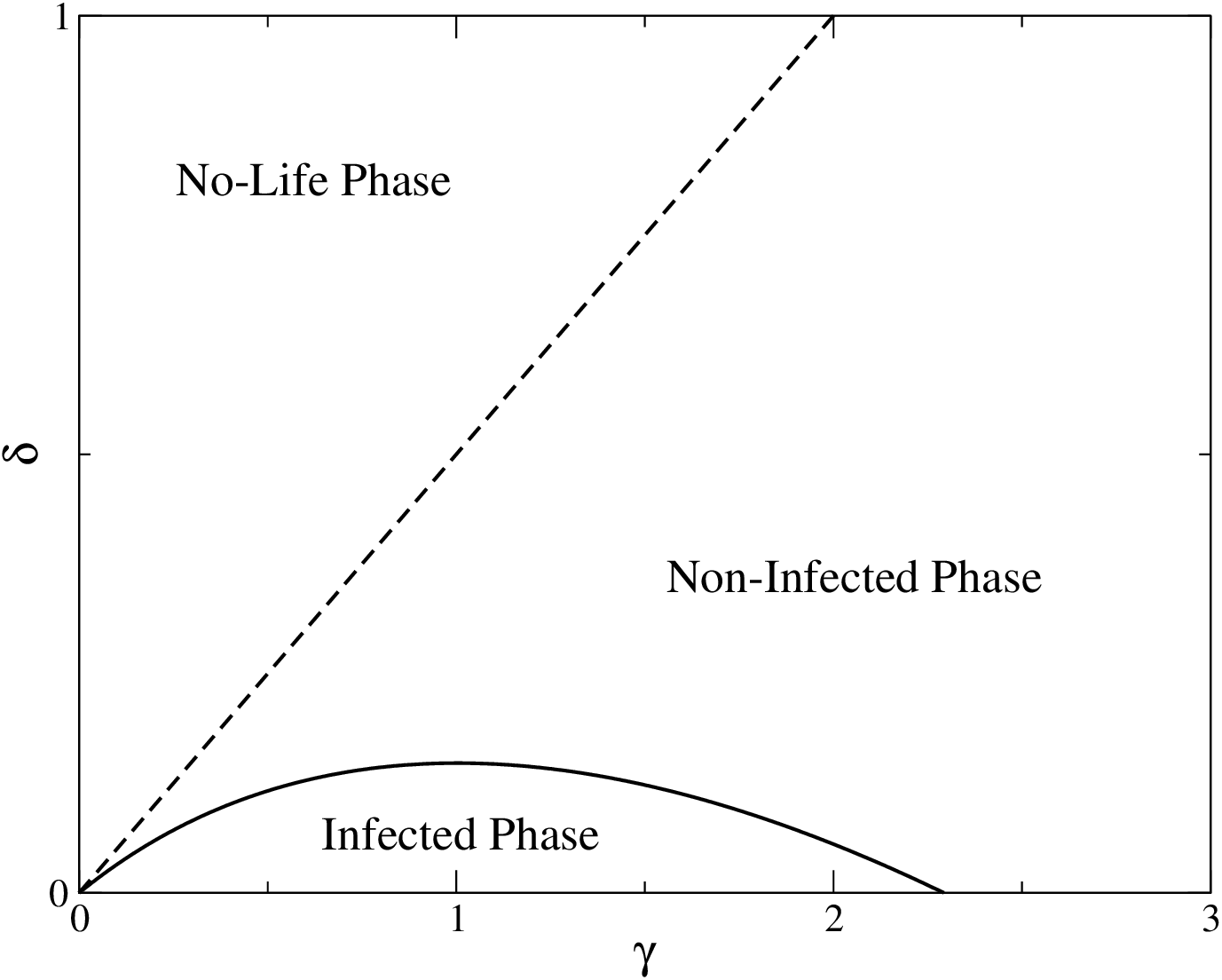}}
\caption{$\delta-\gamma$ phase plot for the mean field equations
(\ref{meanfieldeq}). The straight line is given by
$\delta=\gamma(\beta-1)$ while the curve separating the region of
infection from the region without infection is obtained by finding the
implicit function defined by setting the quantity $\mathcal{E}(\delta,\gamma)$
defined in the text equal to zero. Parameter $\beta$ and $a$ are chosen to be 1.5
and 5 (arbitrary units), respectively.}
\label{dgmeanfield}
\end{figure}
\begin{figure}[tbh]
\centering \resizebox{\columnwidth}{!}{\includegraphics{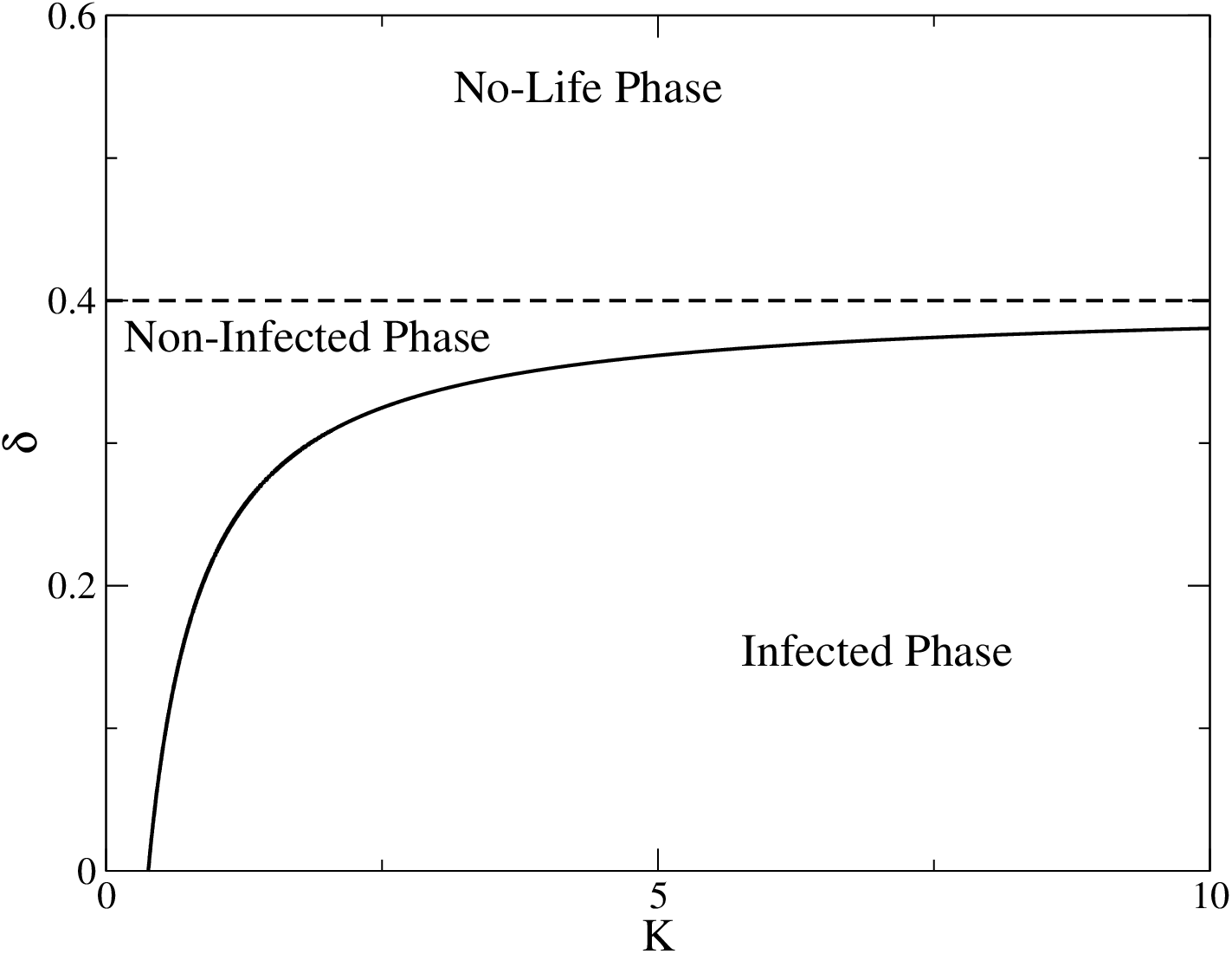}}
\caption{Mean field phase plot in the $\delta-K$ plane.
The horizontal line is given by $\delta=\gamma(\beta-1)$ while the curve separating
the infected from the susceptible phase is simply the graph of Eq. (\ref{kcrit}).
The other parameters are: $\beta=1.4$, $\gamma=1$ and $a=10$ in arbitrary units. }
\label{dKmeanfield}
\end{figure}

The analytical derivation of the steady state mean field allows us to
generate the phase plots of the system and compare them with those from the
simulation. The $\delta -\gamma $ phase plane is presented in Fig. \ref{dgmeanfield}.
It shows when infection is present and when a non-zero population exists.
The straight line is $\delta =\gamma (\beta -1)$ while the curve
limiting the region of infection is obtained by finding the implicitly
defined function $\mathcal{E}(\delta,\gamma)=0$.
The reasons for the presence of a maximum in the $\mathcal{E}(\delta ,\gamma )=0$
curve can be explained as follows. When $\gamma =0$, juveniles cannot
become adults and new juveniles cannot be born. The system is thus
driven to extinction because of predation and competition.
Alternatively, it is easy to see that the only steady state solutions in
Eq. (\ref{sumeq}), when $\gamma=0$, are the trivial case $\mathcal{A}=\mathcal{B}=0$.
An increase in $\gamma $ from zero helps
infection since adults are necessary to generate more juveniles.
The latter are primarily responsible for spreading the infection.
A non-zero $\gamma$ may thus give rise to infection if $\delta$ is sufficiently
small as can be seen by the sublinear increase of the
$\mathcal{E}(\delta,\gamma)=0$ curve in Fig. \ref{dgmeanfield}.
However, beyond a certain value of $\gamma$ infection is hampered:
juveniles are converted too fast
into adults and do not have the time to meet other individuals and spread
the infection. This effect is the more pronounced the larger
the predator pressure. This explains the curvature of the
$\mathcal{E}(\delta,\gamma)=0$ curve beyond its maximum. 
Simulations do not possess
this transition since clusters of adults are generated by increasing
$P_{g}$ in the system. Infection is actually helped by increasing $P_{g}$
beyond a certain value.

Differences between the simulation and the mean field results are also
observed when comparing the $\delta -K$ phase plot in Fig.
\ref{dKmeanfield} with the corresponding simulation phase plot in Fig.
\ref{PpPksimul}. Qualitatively different curves separate the regions of
non-zero steady state populations from regimes where no population
persists at long times. In the mean field the no-life phase is independent of $K$ while
it strongly depends on $P_{K}$ in Fig. \ref{PpPksimul}. This difference
is understandable since in the simulation the death by predators and the
death by competition act together to drive the system to extinction,
while in the mean field, only if
$\delta >\gamma (\beta -1)$ or if $K=0$, the steady state population is zero.
Such differences in extinction behaviour in the predictions of mean field
versus more accurate descriptions can arise because discreteness and finiteness
of numbers of mice are neglected in the mean field analysis. Comments on these
differences have appeared in earlier literature~\citep{aguirre02,escudero04}.

The other important qualitative difference between the two levels of
description of our model can be seen by noticing that
\begin{equation}
K_{c}\approx \frac{1}{a}\frac{f(\beta ,\gamma )}{(\beta -1)\gamma -\delta }, \\ \nonumber
\end{equation}
as $\delta$ approaches $\gamma (\beta -1)$ from below. Here,
$f(\beta ,\gamma )=(2\gamma
)^{-1}(1+\gamma )(2+\gamma )(1+\beta \gamma )\left\{ \left[ 1+4\beta
\gamma ^{2}/(2+\gamma )^{2}\right] ^{1/2}-1\right\} $. This means that the
line that separates the infected from the susceptible phase in Fig.
\ref{dKmeanfield} converges at large $K$ to the line that separates the
susceptible phase from the phase with zero population. Therefore,
as the population grows (linearly) with $K$, an infected phase will always appear
beyond a certain value of
the environment parameter. In the
mean field a region in parameter space where the population,
upon increase of $K$, does not become infected
exists only in the trivial case corresponding to $a=0$.

\section{Conclusions}
\label{conclusions}

The analysis of the effect of predators of juvenile rodents presented in this paper
completes our preliminary studies of the so-called LS model of stationary adult rodents and itinerant
juveniles introduced earlier~\citep{liqsolid}. We have found that a non-infected phase
emerges as a transcritical bifurcation as function of the predation `pressure'.
The LS model owes its existence to a detailed study of field observations on rodents carried out
recently~\citep{panama,sevilleta} on \emph{Zygodontomys brevicauda} in Panama and
on \emph{Peromyscus maniculatus}~\citep{stickelbook} in New Mexico.
That study necessitated a generalization of a previous model~\citep{AK} into the LS
model~\citep{liqsolid} to incorporate observed home ranges~\citep{jtbhomerange}.

A realistic modeling of the Hantavirus epidemic needs to represent the fact that
an adult mouse is less prone to predation in comparison to a juvenile mouse.
Familiarity of the area inside a home range provides a resident mouse with
higher security and shelter from intruders, minimizing the chance of its being
killed by predators.
A non-trivial utilitarian consequence of our present analysis is the possibility of buffering and even
eliminating infection without killing off the mouse population.
Ongoing work in our group is focussed on kinetic level investigations as well as traveling
wave studies in these systems.

\begin{ack}
We acknowledge helpful conversations with David MacInnis, Guillermo
Abramson, Marcelo Kuperman, Terry Yates and Robert Parmenter.
This work was supported in part by the NSF
under grant no.
INT-0336343, by NSF/NIH Ecology of Infectious Diseases under grant no.
EF-0326757, and by DARPA under grant no.
DARPA-N00014-03-1-0900. We acknowledge the Center for High Performance
Computing, UNM, for making available to us their computing resources.
\end{ack}

\end{document}